\documentclass[aps,pra,showpacs,twocolumn,amsfonts,amssymb,amsmath]{revtex4} 
\usepackage{graphicx} 
\usepackage{array}
\usepackage{SIunits}
\usepackage{color} 
\def\T2{$\mathrm{T_2}$}
\def\Pr3{$\mathrm{Pr^{3+}}$}

\def\PrYSO{Pr$^{3+}$:Y$_2$SiO$_5\,$}
\def\mrm#1{$\mathrm{#1}$}

\def\eq#1#2{\begin{equation}\label{#1}#2\end{equation}}
\def\al{$\alpha L\,$}

\begin{document}

\title{Experimental superradiance and slow light effects for quantum memories}

\author{A.~Walther, A.~Amari, S.~Kr\"{o}ll} \affiliation{Department of Physics, Lund Institute of Technology, P.O.~Box 118, SE-22100 Lund, Sweden}

\author{A.~Kalachev}\affiliation{Zavoisky
Physical-Technical Institute of the Russian Academy of Sciences,
Sibirsky Trakt 10/7, Kazan, 420029, Russia}

\date{\today }

\begin{abstract}
The effects of high optical depth phenomena, such as superradiance, are investigated in potential quantum memory materials. The results may have relevance for several schemes, including CRIB, AFC and EIT-based quantum memories, which are based on using ensembles as storage media. It is shown that strong superradiant effects, manifested as decay rates larger than $1/T_2^\ast$, are present even for moderate values of \al $\leq 5$, and increases as a function of \al. For even higher \al, effects like off-resonant slow light is demonstrated and discussed, and finally, the efficiency of time-reversed optimized input pulses are tested. A maximum retrieval efficiency of $\sim$~20\% is reached, and agreement with the theoretically expected result is discussed.
\end{abstract}


\maketitle

\section{Introduction}
With a rapidly growing need for quantum communication networks for applications like quantum cryptography or quantum computing, it becomes important to be able to send single quantum information packets reliably, and across great distances. Photons are good candidates for such long distance carriers, but with the best currently achievable fibers, losses in the order of 0.2 dB/km will put an effective limit on the maximum distance of about 50 - 100 km \cite{Tittel2008}. One way to overcome this weakness is by using quantum repeaters, which were introduced in 1998 by Briegel \emph{et al} \cite{Briegel1998}. In this scheme the total distance is divided into subparts where entanglement between nodes can be created and stored individually for each segment. This greatly increases the efficiency. For high repetition rates, the scheme relies on quantum memories which can store and recall single photons with high efficiency as well as high fidelity. There have been several suggestions for the implementation of such quantum memories, including EIT/Dark state polaritons \cite{Fleischhauer2002}, off-resonant Raman interactions \cite{Duan2001}, Controlled reversible inhomogeneous broadening (CRIB) \cite{Moiseev2001,Nilsson2005a,Kraus2006} and, most recently, Atomic frequency combs (AFC) \cite{Afzelius2008,Riedmatten2008}.

In all of the schemes mentioned above, high optical depth is crucial in order to obtain high efficiency for the storage and recall process. In this Paper, consequences of being in the high \al regime, such as superradiance \cite{Dicke1954} and slow light effects, are investigated. Superradiance is a collective quantum emission effect and is one of the major mechanisms enabling quantum memories, since the increased emission rates it gives are essential for separating the recalled pulse from spontaneous emission. For high optical depth situations however, there are also additional effects, like more rapid loss of the stored energy with increasing \al. This may lead to lower expected maximum recall efficiencies, but can also help to facilitate faster multimode sequences, since the superradiance effect causes shorter pulses to become more efficient. Using a system that allow the optical depth of the ensemble to be changed easily, we have here investigated the duration of the superradiant emission as a function of \al.

A very useful choice of qubit for long distance quantum communication is the time bin qubit \cite{Brendel1999}, were the superposition consists of a pulse being either in an earlier 'time bin' or in a later one. In this scheme, using attenuated pulses, one has the freedom to choose the distribution of the input photon (pulse shape). Theoretical work has been done to find shapes that optimize efficiency \cite{Gorshkov2007}, or signal-to-noise ratio \cite{Kalachev2007}, even well inside the superradiant regime. Here, we still have sufficiently low \al that a simple time-reversed replica of the exponential decay is a good choice with regards to both optimization criteria, and for such input pulses, the efficiency as a function of \al was investigated, and compared with the theoretically expected efficiencies.

Both the CRIB and the AFC protocol are aimed at utilizing the natural inhomogeneous structure found e.g. in rare-earth-ion-doped crystals. Such crystals have received a lot of attention in the past decade as promising candidates for quantum computing \cite{Longdell2004a,Rippe2008} and, in particular, for quantum memories \cite{Tittel2008}. The main advantage of these solid state materials is a very long coherence time for both the optical states and the spin states. In particular, an optical state coherence time of several ms has been measured in Erbium \cite{Sun2002} and coherence times of up to 30 s have been measured for the hyperfine states, using decoupling sequences (bang-bang pulses) \cite{Fraval2005}. In the experiments done for this Paper we employ advanced optical pumping techniques in order to tailor specially prepared spectral structures in a part of the inhomogeneously broadened profile in a \PrYSO crystal.

The Paper is organized as follows. In Section~\ref{sec:theory} a brief description of the theory for superradiant forward scattering is presented. In particular, equations describing superradiant decay time and superradiant emission efficiency as a function of \al is presented. In Section~\ref{sec:setup} the experimental setup and preparation of the spectral structures for the experiment is described. In Section~\ref{sec:results} the superradiance and the slow light data is presented together with a brief analysis of the implications to the current quantum state storage protocols, and in particular the AFC protocol. Finally, the Paper is concluded in Section~\ref{sec:conclusion}.

\section{Superradiant forward scattering}
\label{sec:theory}
The phenomenon of optical superradiance (SR), which is a vivid manifestation of constructive interference in the process of collective spontaneous emission of photons by a system of initially excited particles, was theoretically predicted by Robert Dicke in 1954 \cite{D_1954}. Most studies of optical superradiance (see reviews \cite{AEI_1980,GH_1982,BEMST_1996}) are devoted mainly to the particular case called superfluorescence (SF) \cite{BL_1975}. In this case, at the initial moment of time, a pumping pulse inverts the atomic system to an excited state, so that the mean value of the dipole moment of each atom becomes equal to zero. SF occurs owing to the self-inducing of correlations between the electric dipole moments of optical atomic transitions through the common emission field. As a result a system of $N$ inverted atoms can spontaneously return to the ground state for a time inversely proportional to the number of atoms, emitting a coherent light pulse whose intensity is proportional to $N^2$. SF is a striking example of self-organization, when the coherence of atomic states forms from an initially incoherent state. In an extended atomic system this process corresponds to macroscopic polarization wave creation, which is only possible in an optically thick resonant medium.

In a more general case, an excited atomic system may have a nonzero macroscopic dipole moment at the starting time. This can result from the effect of a coherent excitation pulse, which transfers the system directly from the ground state into the superradiant state thereby avoiding the self-organization step. The excitation pulse may here have a small pulse-area, and will not undergo amplification, since no atoms are excited before the pulse. Nevertheless, the collective nature of the spontaneous emission reveals itself in that the decay time decreases with the optical thickness (speed-up) and that the intensity modulated in time (ringing). Both effects can be alternatively viewed as the features of coherent propagation of the small excitation pulse in the optical thick resonant medium. This process is usually referred to as superradiant forward scattering (SFS) especially in $\gamma$-ray optics (see, for example, the review in \cite{S_1999}). The corresponding theory has long been known \cite{BC_1969,C_1970,KAK_1979}. Nevertheless, some new aspects have appeared recently in the connection with the optimization of light-matter interaction motivated by the development of optical quantum memories \cite{GAFSL_2006,K_2007}. In terms of SFS, this means optimization of the excitation pulse aimed at obtaining the maximum scattered energy (efficiency) or peak amplitude (signal-to-noise ratio). These criteria lead to slightly different results, but in any case, the optimal shape of the excitation pulse proves to be a time reversed replica of the emitted pulse. If the optical thickness is not far above unity, the ringing may be neglected and the optimal shape can be approximated by the reversed exponential decay. Let us now consider the basic features of SFS in this case.

Consider an extended system of identical two-level atoms forming a resonant medium of length $L$ and arbitrary large cross section. We assume that the atoms are not moving as impurities embedded in a solid state material. We are interested in the interaction of the atomic system with a resonant coherent light pulse and treat the problem in a one-dimensional approximation. This is a good propagation model provided that the Fresnel number of the interaction volume  $F=S(L\lambda)^{-1}\gg 1$, where $S$ is the cross section of the light beam, and $\lambda$ is the wavelength of the resonant transition. Besides, we restrict our attention to propagation of small-area coherent pulses through the medium. In this case, the resonant medium may be considered as a linear system and the formal solution of the pulse-propagation problem is given by the convolution
\begin{equation}\label{Convolution}
F_\text{out}(t)=\int_{-\infty}^{+\infty}d\tau F_\text{in}(\tau)H(t-\tau),
\end{equation}
where $F_\text{in}(t)$ and $F_\text{out}(t)$ are the envelopes of the input and output pulses, respectively, and $H(t)$ is an impulse-response function \cite{C_1970,KAK_1979},
\begin{equation}\label{Impulse Response}
H(t)=\delta(t)-b\frac{J_1(2\sqrt{bt})}{\sqrt{bt}}\,\theta(t)\,e^{-t/T_2}.
\end{equation}
Here $\delta(t)$ is the Dirac delta function, $\theta(t)$ is equal to 0 for $t<0$, 1 for $t>0$, and $1/2$ for $t=0$, $J_n(x)$ is the Bessel function of the first kind, $T_2$ is the phase relaxation time for the atomic transition, $b=\alpha L/2T_2$ is the thickness parameter, and $\alpha$ is a resonant absorption coefficient. It is assumed that the absorption line broadening is Lorentzian in shape with the linewidth $\Gamma=1/\pi T_2$. The quantity $T_R=1/b$ is referred to as superradiant life-time.

Let the shape of the input pulse be
\begin{equation}\label{F_in}
F_\text{in}(t)=\sqrt{{2}/{T}}\,e^{t/T}\theta(-t).
\end{equation}
The pulse is normalized so that $\int_{-\infty}^{\infty}|F_\text{in}(\tau)|^2\,d\tau=1$.
Substituting (\ref{F_in}) into (\ref{Convolution}) and taking into account (\ref{Impulse Response}) we obtain
\begin{equation}\label{F_out}
F_\text{out}(t)=F_\text{in}(t)-\Phi(0)\,e^{t/T}\,\theta(-t)-\Phi(t)\,e^{t/T}\,\theta(t),
\end{equation}
where
\begin{equation}
\Phi(t)=\sqrt{\frac{2}{T}}b \int_{t}^{\infty} d\tau \frac{J_1(2\sqrt{bt})}{\sqrt{bt}}\,e^{-\tau(1/T_2+1/T)}.
\end{equation}

To begin with, we note that
\begin{equation}\label{F0}
\Phi(0)=\sqrt{\frac{2}{T}}\left(1-e^{-b(1/T+1/T_2)^{-1}}\right).
\end{equation}
This quantity determines the amplitude of the decay pulse ($F_\text{out}(t)$ in the limit $t\to +0$) as a function of optical thickness, phase relaxation time and duration of the excitation pulse. It follows from Eqs.~(\ref{F_out}) and (\ref{F0}) that in the case of high optical thickness ($\alpha L\to\infty$), the scattered signal appears only after termination of the excitation pulse having the maximum possible amplitude $\sqrt{2/T}$.

It is now instructive to consider several cases when Eq.~(\ref{F_out}) simplifies and allows the integration.

(a) The resonant medium is optically thin, i.e. $\alpha L\ll 1$. Then $J_1(2\sqrt{b\tau})/\sqrt{b\tau}\approx 1$ and
\begin{multline}\label{thin_medium}
F_\text{out}(t)=\sqrt{\frac{2}{T}}\left(1-b\frac{TT_2}{T+T_2}\right)\,e^{t/T}\,\theta(-t)\\
-\sqrt{\frac{2}{T}}b\frac{TT_2}{T+T_2}\,e^{-t/T_2}\,\theta(t).
\end{multline}
The decay process after the excitation pulse is exponential and is nothing more than the FID. The maximum amplitude of the FID is achieved when $T=T_2$.

(b) The duration of the excitation pulse $T$ is much shorter than the superradiant life-time $T_R$ and phase relaxation time $T_2$. Then
\begin{multline}\label{short_excitation}
F_\text{out}(t)=\sqrt{\frac{2}{T}}(1-bT)\,e^{t/T}\,\theta(-t)\\-\sqrt{\frac{2}{T}}bT\frac{J_1(2\sqrt{bt})}{\sqrt{bt}}\,e^{-t/T_2}\,\theta(t).
\end{multline}
The decay process after the short pulse excitation reproduces the impulse-response function, as expected. If the resonant medium is optically thin, Eq.~(\ref{short_excitation}) reduces to Eq.~(\ref{thin_medium}) with $T\ll T_2$. In the opposite case, when $\alpha L\gg 1$ the decay process is modified by the Bessel function, i.e. it is accelerated and modulated in time.

(c) Superradiant life-time $T_R$ is much shorter than the duration of excitation pulse $T$ and phase relaxation time $T_2$. In this case, we perform the integration using
\begin{equation}
\frac{\partial J_0(2\sqrt{b\tau})}{\partial\tau}=-b\frac{J_1(2\sqrt{bt})}{\sqrt{bt}}
\end{equation}
and obtain
\begin{equation}
F_\text{out}(t)=-\sqrt{\frac{2}{T}}J_0(2\sqrt{bt})\,\theta(t),
\end{equation}
provided that $bT$ and $bT_2\gg 1$. The decay process undergoes similar modifications, but oscillations are much more prominent in this case.

In general, despite the oscillations it is possible to fit the decay curve by exponential
\begin{equation}\label{model}
|F_\text{out}(t)|^2= I(0)\exp(-t/T_\text{dec}),
\end{equation}
where
\begin{equation}\label{decay_time}
\frac{1}{T_\text{dec}}=\frac{2}{T_2}+\frac{\alpha L}{2T_2}x
\end{equation}
is the effective decay time. The parameters $x$ and $I(0)$ are functions of optical thickness, phase relaxation time and duration of the excitation pulse, which can be calculated numerically from Eq.~(\ref{F_out}). In doing so, it is more convenient to rewrite Eq.~(\ref{model}) in the following form
\begin{equation}\label{model2}
\int_0^t d\tau |F_\text{out}(\tau)|^2= \mathcal{E}\left(1-e^{-t/T_\text{dec}}\right),
\end{equation}
where
\begin{equation}\label{eq:eff}
\mathcal{E}=\int_0^\infty d\tau |F_\text{out}(\tau)|^2=I(0) T_\text{dec}
\end{equation}
is the efficiency of SFS. In the cases (a) and (b) considered above, we obtain $I(0)\approx |\Phi(0)|^2$ and $x\sim 1$, while in the case (c), $I(0)\ll |\Phi(0)|^2$ and $x\ll 1$ due to the weakly decaying oscillations. Specifically, in the case of short excitation pulses, $\mathcal{E}=2Tb[1-I_0(bT_2)\,e^{-bT_2}-I_1(bT_2)\,e^{-bT_2}]$, where $I_n$ is the modified Bessel function of the first kind. Therefore, taking $I(0)=|\Phi(0)|^2$, we obtain $x\approx 1+\alpha L/24$ for $\alpha L \lesssim 1$ and $x\approx 1+2/\sqrt{\pi\alpha L}$ for $\alpha L\gg 1$. Finally, under conditions of the present experiment described below, numerics show that $x\approx 1+0.055\alpha L$ for $T=T_2/2$ and $0.5 \leq \alpha L \leq 5$, provided that $I(0)=|\Phi(0)|^2$.

\section{Experimental preparations}
\label{sec:setup}
\subsection{Setup}
The experiments were performed on the $^3$H$_4-^1$D$_2$ transition of praseodymium ions doped into yttrium silicate (\PrYSO) in site 1, absorbing at 605.82 nm. The sample was 20 mm long in the direction of light propagation and had a width and height of 10 mm, respectively, with two polished curved end-faces. The crystal was immersed in liquid helium and kept at a temperature of 2.1 K in an Oxford Instruments Spectromag cryostat, and had a \Pr3 concentration of 0.05\%.

A ring dye laser (Coherent699-21) using Rhodamine 6G pumped by Nd:YVO$_4$ laser (Coherent Verdi) was used to give 600mW output power at $\lambda=605.82$ nm. The Dye laser is stabilized against a spectral hole in a \PrYSO crystal, yielding a coherence time $>100$ \micro s and a frequency drift $<1$ kHz/s \cite{Julsgaard2007}. In order to create the desired pulses shapes a 200 MHz acousto-optic modulator (AOM) with a bandwidth of 100 MHz was used. It was aligned in a double pass configuration in order to avoid directional movement of the diffracted beam as a function of frequency. The pulses to the AOM were generated by a 1 GS/s arbitrary waveform generator (Tektronix AWG520), which allowed direct control of pulse amplitude, phase, and frequency. After the AOM, the light passed through a single mode optical fiber to clean up the spatial mode. A beam sampler removed a small percent of the light before the sample to be used as a reference beam. The rest of the beam passed through a $\lambda$/2 plate, where the light polarization could be adjusted to match the crystal orientation, and was focused onto the center of the crystal. The diameter of the focus was 100 \micro m throughout the sample, which gave Rabi frequencies of a maximum 2 MHz for the active transition.

\subsection{Preparing spectral structures}
\begin{figure}[h]
        \includegraphics[width=8.5cm,height=4cm]{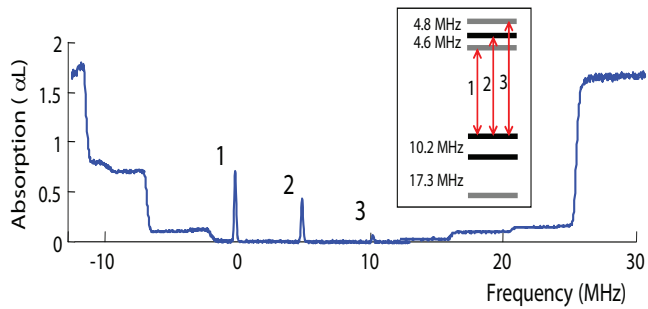}
    \caption{(color online) Created spectral structure containing three peaks corresponding to all possible transitions from one of the hyperfine ground states to the excited hyperfine states, inside an empty pit somewhere on the inhomogeneous profile. The inset shows the energy level diagram for \Pr3.}
    \label{Peaks}
\end{figure}
All experiments were done on an ensemble containing in the order of $10^8-10^{10}$ \Pr3 ions (depending on \al), which had been spectrally isolated inside the absorption profile. The full inhomogeneous profile of \PrYSO is about 5 GHz wide. Optical pumping techniques were then used to completely empty an 18 MHz wide interval, henceforth known as a spectral \emph{pit} (see also descriptions in Refs.~\cite{Rippe2005,Rippe2008}). The optical pumping procedure involved moving all ions over to the lowest state, seen in the \Pr3 energy level diagram in the inset of Fig.~\ref{Peaks}. From that state, a $\sim$120 kHz narrow interval of ions was coherently burnt back to the upper ground state, into the otherwise empty pit. The resulting spectral structure is shown in Fig.~\ref{Peaks}, where the three peaks correspond to the same ions contributing to the absorption from one of the hyperfine ground states to each of the three possible excited states. The reason for the different heights of the peaks is different oscillator strengths for the transitions. During the experiment, the ion ensemble was sometimes prepared in other hyperfine ground states as well, but the technique for creating those are similar to the one described.

The spectral structures were measured by scanning the light frequency and recording the intensities of both the transmitted and the reference beams. Both beams were detected by Thorlabs PDB150A detectors and the signals from the detectors were divided to reduce the effect of laser amplitude fluctuations. The intensity of the probe pulses were small enough not to affect the created spectral structures during the readout process.

\section{Results and discussion}
\label{sec:results}
The discussions will be divided into two parts. First, the superradiance results will be presented, and second, the results related to the accompanying slow light effects.

\subsection{Superradiance measurements}
\begin{figure}[h]
    \includegraphics[width=8.5cm,height=4cm]{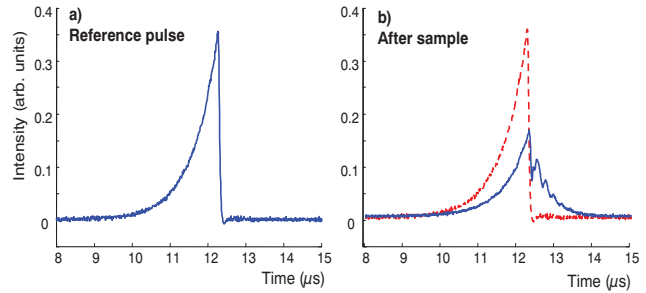}
    \caption{(color online) a) Input pulse shape corresponding to a reversed exponential decay (see Eq.~\ref{F_in}). b) Output pulse, after it has passed through a sample with $\alpha L \approx 4.5$. The pulse is partially absorbed and immediately reemitted in the form of superradiant decay. The dashed red line in part b) is the input pulse as a reference.}
    \label{SR_pulses}
\end{figure}
The optimized reversed pulses given in Eq.~\ref{F_in}, were used to test the duration of the superradiant decay. They were implemented experimentally using the AOM system, and an example of such a pulse is shown in Figure~\ref{SR_pulses}. Part a) displays the pulse on the reference detector that does not pass through the sample, and part b) shows the same pulse on the detector after the sample where a spectral peak has been prepared. Most of the pulse will get absorbed as it passes through the sample, and will then be immediately reemitted in the form of a superradiant decay. This phenomenon is similar to that of free induction decay (FID), but in the case of FID, the duration of the decay is determined by the inverse of the inhomogeneous width of the peak, i.e. $\tau_{FID} = T_2^\ast = 1/\pi\Gamma_{peak}$. As long as the ions on the different resonance frequencies are still in phase, they are emitting with an intensity that is proportional to the square of the number of ions involved, i.e. $I_{coherent} \sim N^2$. After the effective coherence time however, they will have dephased, and will thus only emit light incoherently with $I_{incoherent} \sim N$. With roughly $10^9$ ions in a peak the coherent effect is very large, and this is the reason why the emission is limited to times shorter than \mrm{T_2^\ast}.

\begin{figure}[h]
    \includegraphics[width=8.5cm,height=6.61cm]{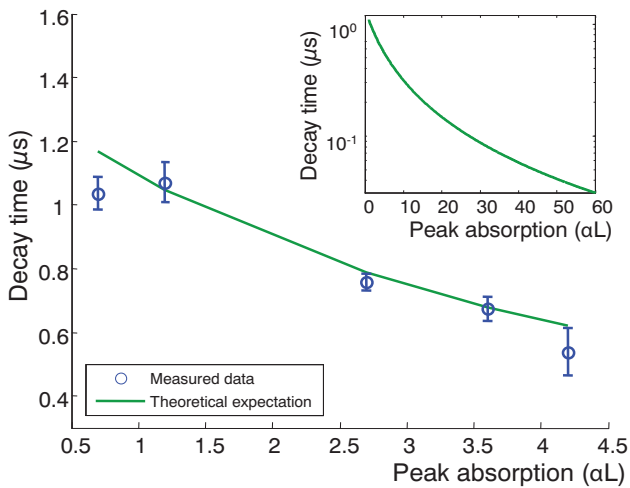}
    \caption{(color online) Duration of the superradiant decay plotted as a function of peak \al. The inset shows an extension of the theoretical line, i.e. what can be expected for higher \al.}
    \label{SR_duration}
\end{figure}

In the regime of high optical depth however, $N$ is so large that, the intensity becomes large enough to cause a substantial amount of the ions to become deexcited in a time shorter than \mrm{T_2^\ast}. This causes the duration of the superradiant decay to become proportional to the inverse of the number of ions involved, as can be inferred from Eq.~\ref{decay_time}. In this regime we have studied the rate of the superradiant decay for the reversed optimal pulses, resonant with peaks of different \al. For all cases, the same pit structure was created, but the Rabi frequency of the coherent pulses that created the peak was varied, which in turn yielded peaks inside the pit of correspondingly varying \al. Figure~\ref{SR_duration} shows the duration of the superradiant decay as a function of \al. All experimental points in the graph are results from cases where the decay duration was measured on a single shot basis, and then averaged with other shots of the same peak \al. The (blue) circles corresponds to the experimental data and the solid (green) line is the expected decay times according to Eq.~\ref{decay_time}. As can be seen, the decay time is decreasing with \al and the experimental data come very close to the theoretical predictions. The remaining discrepancy can partially be explained by laser phase jitter, in particular during the peak creation procedure. If peaks with varying width is created, the superradiant decay will also vary correspondingly, since the decay time is proportional to \mrm{T_2^\ast} (cf. Eq.~\ref{decay_time}).

The theoretical line has been extrapolated up to higher \al, displayed in the inset to Fig.~\ref{SR_duration}. This was done in order to estimate what effects would be obtained for the optical depths needed to provide high efficiency for the various quantum memory protocols. In AFC for example, the atomic comb structure has a certain peak width, $\gamma$, and a peak separation of $\Delta$, which allows the finesse to be defined, $F = \Delta/\gamma$. In Ref.~\cite{Afzelius2008}, it is stated that the AFC memory could obtain an efficiency above 90\% for a reasonable comb finesse of 10, when the optical depth is in the order of $\alpha L \approx 40$. As a comparison, from Eq.~\ref{decay_time}, taking $x=1$, we get that for a typical currently used value of $\alpha L = 2$, the decay time can be expressed $T_{dec} = T_2^\ast/3$, while for $\alpha L = 40$, we instead get $T_{dec} \approx T_2^\ast/22$. Thus, by increasing $\alpha L$ a factor of 20 we decrease the time it takes for the superradiant decay to drain the excited state by a factor of $\sim 7$. In the AFC protocol, the time between the absorption event and the reemission is $T_{recall} = 1/\Delta$, following the first occasion where all peaks have rephased. However, some light will be collectively emitted even from a single peak, on a time scale given by the inverse of frequency width of each peak, $T_{single} = 1/\pi\gamma$. With a good finesse, the peak width is much narrower than the peak separation, and thus $T_{single} \gg T_{recall}$, which means little light is lost due to emission from the single peaks. When moving into the superradiant regime however, the decay from each single peak, instead becomes proportional to number of atoms in the peak, and has a duration which is decreasing with \al. Taking the values used in the example above, increasing \al from 2 to 40 decreases $T_{single}$ by a factor of 7 which can even cause it to become shorter than the total rephasing time, $\left(T_{single}\right)_{SR} < T_{recall}$. In general, given Eq.~\ref{decay_time}, the duration of the superradiant decay due to single peak collective emission, can be expressed as
\eq{SR_AFC}{
  \left(T_{single}\right)_{SR} = \frac{F}{\pi\left(2+\frac{\alpha L}{2}x\right)}T_{recall},
}
where $x\approx 1+2/\sqrt{\pi \alpha L}$ for high \al and short excitation pulses. We see that in the example case of $\alpha L = 40$ and $F = 10$, the single peak decay time is around $\left(T_{single}\right)_{SR} \approx 0.12 \cdot T_{recall}$, which is a quite substantial change.

\begin{figure}[h]
    \includegraphics[width=8.5cm,height=4.6cm]{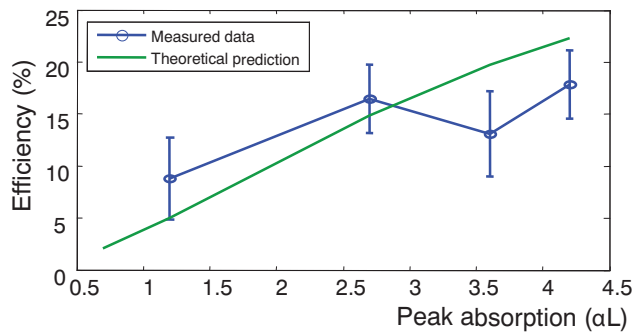}
    \caption{(color online) Efficiency of the superradiant decay, in terms of how much energy is in the decay compared to the total energy of the input pulse. Blue circles is the experimentally measured, and the green solid line is the theoretical predictions given by Eq.~\ref{eq:eff}.}
    \label{Efficiency}
\end{figure}

The efficiency is defined as the total energy found in the superradiant decay compared to the total energy of the incoming pulse, and is plotted in Figure~\ref{Efficiency} as a function of \al. The experimental data is again the (blue) circles, which in general are reasonably close to the theoretical predictions of Eq.~\ref{eq:eff}, represented by the (green) solid line. Each circle in the figure corresponds to an average of 10 single shot values and the remaining discrepancies are contributed to laser frequency fluctuations. In connection to the discussion for AFC above, we also note that the efficiency grows with \al. This means that not only is the decay due to SFS faster and faster with increasing \al, it also contains more and more of the stored light, which could eventually lead to this process being a substantial loss mechanism.

\subsection{Slow light effects}
\begin{figure}[h]
    \includegraphics[width=8.5cm,height=9.4cm]{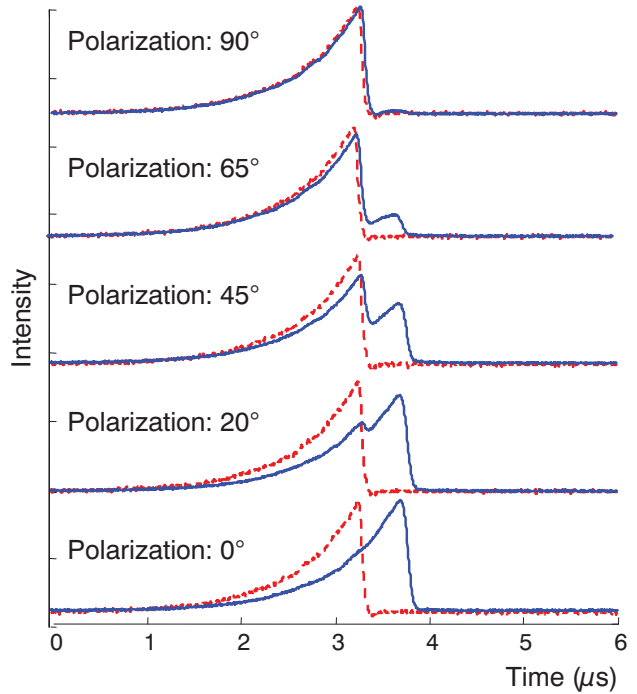}
    \caption{(color online) The polarization angle is given relative to the atomic dipole moment orientation. At 90$^\circ$, displayed at the top, the pulse polarization is perpendicular to the atomic dipole moment, and no slow light effect is seen. Tilted 45$^\circ$, each photon in the pulse has a 50\% chance to be parallel with the atoms in which case it gets delayed, otherwise not. When the pulse polarization is fully parallel with the atoms, all of the transmitted pulse is delayed. The intermediate polarization steps at 20 and 65$^\circ$ further verify the behavior that each photon has a chance to end up in one of the two time bins, with a respective probability given by the polarization angle.}
    \label{Slow_light}
\end{figure}

Slow light effects (see e.g. \cite{Milonni2002}) are mostly observed and investigated in experimental situations where simultaneous electro-magnetic fields resonant with transitions in a material are used to modify the frequency dependence of the index of a refraction in a material, such that the group velocity of at least one of the interacting fields is changed significantly, e.g. \cite{Hau1999,Turukhin2002}. However, it was also proposed \cite{Shakhmuratov2005,Rebane2007} that absorption profiles with basically arbitrary spectral structures could be carved out in persistent hole-burning materials and subsequent pulses propagating through such spectrally crafted material could then experience significant light group velocity slow down. Rare-earth-ion-doped crystals at cryogenic temperatures can be particularly suitable for such experiments since spectral structures created in these materials may persist for days \cite{Koenz2003}. Results along these lines were also demonstrated by Sellars and Manson, where the group velocity was reduced to 60 m/s by creating a 2 kHz wide spectral hole in a Eu:Y$_2$SiO$_5$ crystal \cite{Sellars2003}. In our case the zero absorption region (spectral hole) is about 18 MHz, but as can be seen in Fig.~\ref{Peaks}, the absorption is actually low within a 30 MHz interval from about -5 MHz to +25 MHz. When the inhomogeneous line width, as in this case, is much larger than the spectral hole line width (FWHM), $\Gamma$, and the laser line width is much narrower than the spectral hole, Shakhmuratov et al. \cite{Shakhmuratov2005} showed that the group velocity, $v_g$, for a pulse propagating within the spectral hole is basically given by $v_g=\pi \cdot \Gamma/\alpha$, where $\alpha$ is the absorption coefficient outside the spectral hole. For our crystal of length, L=2 cm, we estimate $60<\alpha L<80$ at the center of the inhomogeneous profile. With  $\Gamma=30$ MHz and $\alpha L=70$, we obtain $v_g \approx 27000$ m/s. In Fig.~\ref{Slow_light} the dashed (red) curves are the input pulses recorded by the detector in front of the sample and the solid (blue) curves show the pulses after the 2 cm sample. The delay is about 500 ns corresponding to a group velocity of about $40000$ m/s. The discrepancy between the experimentally measured value and the theoretical expectation is thus less than a factor of two, which is quite good considering the simplicity of the formula used. The remaining difference can be explained by the fact that the formula does not take the shape of the hole into account, which in our case is rather special as can be seen from the pit edges visible in Fig.~\ref{Peaks}. When it comes to quantum memory protocols, both CRIB and AFC were designed to work in spectral pits such as those created here, and these protocols also require the high \al regime to achieve optimum efficiency. Slow light effects inside a pit with high \al, may thus limit the access times of such quantum memories. Although, for a scheme with good multi mode capacity, such as AFC, the effects of long access times can be greatly reduced, since the mode repetition rate is not affected by the slow light.

Interestingly the delay is polarization dependent. Due to the crystal site symmetry and the orientation of the transition dipole moments, rotating the light polarization 90$^\circ$ produces no delay as can be seen in the upper part of Fig.~\ref{Slow_light} and any intermediate polarization produces one delayed and one non-delayed pulse. This is illustrated in the rest of Fig.~\ref{Slow_light}, where the polarization has been successively brought closer towards the direction of maximum absorption, with an increasing proportion of the photons ending up in the delayed part as a consequence. In fact, we are not aware of previous observations of such a polarization dependent slow light effects. For example, a single photon at 45$^\circ$ polarization would generate time-bin qubits where the first and the second probability amplitude distributions are polarized at 90$^\circ$ angle with respect to each other. In essence, the spectral pit here acts like a polarizing time-delay, very similar to the action of a polarizing beam-splitter, something that could potentially have some interesting applications considering the relative simplicity of the creation procedure.

\section{Conclusion}
\label{sec:conclusion}
To summarize, we have investigated the decay immediately following the excitation of an ensemble, with high enough \al to be in the superradiant regime. The excitation pulses used were time-reversed replicas of the exponential decay process. It was found that the duration of the superradiant decay decreased with increasing \al, at the same time as the total energy of the decay pulse increased. A decay duration of 1/3 of what would be expected from free induction decay was measured while the maximum efficiency obtained was approximately 20\%. Implications of these results for quantum memories were discussed and it was found that superradiant decay could be a potentially large loss mechanism in future memory implementations using high \al.

In addition, substantial slow light effects were observed, inside a spectral pit where high \al outside the pit created a large dispersion across the empty region. The slow light effect was found to vary with light polarization. This can be used for creating polarization dependent time-bin qubit states, which could have some interesting applications.

\section*{Acknowledgments}
The authors would like to thank Lars Rippe for insightful comments. This work was supported by the Swedish Research Council, the Knut and Allice Wallenberg Foundation, the European Commission through the integrated project QAP, the RFBR (grant No.~09-02-00206-a) and the Program of the Presidium of RAS "Quantum physics of condensed matter".

\bibliographystyle{prsty}
\bibliography{../bibtex/Ref_lib,PRA1}

\end{document}